\documentclass[runningheads]{llncs}
\usepackage[T1]{fontenc}
\usepackage{graphicx}
\usepackage[export]{adjustbox}
\usepackage{tabularx}
\usepackage{tabularx,makecell,array}

\usepackage{multirow}
\usepackage{subfig}
\usepackage{caption}
\usepackage{cite}
\usepackage{amsmath,amssymb,amsfonts}
\usepackage{algorithmic}
\usepackage{graphicx}
\usepackage{textcomp}
\usepackage[dvipsnames]{xcolor}
\definecolor{red}{rgb}{1.0, 0.3, 0.3}
\definecolor{gold}{rgb}{0.8, 0.6, 0.2}
\definecolor{blue}{rgb}{0.25, 0.5, 1}
\definecolor{brown}{rgb}{0.6, 0.4, 0.2}
\begin{document}
\title{An Energy-Efficient Approximate Posit Multiply–Divide Unit}
%
%
\author{Rishi Thotli\inst{1}
\and
Aditya Anirudh Jonnalagadda\inst{1}\orcidID{0000-0003-3011-1273}  
\and
Rushabh Hulsurkar\inst{2}
\and
Uppugunduru Anil Kumar\inst{3}
\and
Sreehari Veeramachaneni\inst{4}
\and
Syed Ershad Ahmed\inst{1}
\and
John~L.~Gustafson\inst{5}\orcidID{0000-0002-2957-1304}}
\authorrunning{R. Thotli et al.}
%
\institute{
Birla Institute of Technology and Science, Hyderabad, India
\and
Birla Institute of Technology and Science, Pilani, India\\
\and
The ICFAI Foundation for Higher Education, India\\
\and
Sri Sivasubramaniya Nadar College of Engineering, India
\and
Arizona State University, Tempe, Arizona USA\\
\email{rishithotli@outlook.com}\
\email{adi.anirudh1610@gmail.com}\\
\email{jlgusta6@asu.edu}
}

\maketitle              
\begin{abstract}
  In modern computing units, division operations are generally slower than other arithmetic operations and require more resources, such as area and power, than multiplication. To reduce the delay, fast division algorithms use an initial approximation of the reciprocal of the divisor and iteratively approach the correct value, followed by multiplication with the dividend. The hardware architecture and choice of algorithm can significantly alter the overall performance of the division unit. This paper proposes a reduced-accuracy division method for the \textit{posit} number system, which is an alternative to the traditional floating-point system. The proposed design uses a Look-Up Table (LUT) and a single subtraction operation to perform approximate divisor reciprocation by leveraging the mathematical symmetries of the posit number system. The paper also presents a hardware architecture that combines multiplication and division units. The reciprocal calculation has been incorporated into the posit Decoder, a common unit required to perform any hardware operation with posits. Compared to existing hardware implementations of division, the proposed method requires significantly fewer operations at the cost of perfect rounding for division. The proposed architecture was simulated using the Cadence RTL v7.1 E2 compiler at the TSMC 90 nm process node and achieves a Power Delay Product (PDP) reduction of 78.8\% compared to an existing design that performs exact division, while only 46.33\% of the area is required. The experimental results also demonstrate the effectiveness of the proposed system in improving the efficiency of multiplication in posit-based systems.

.
\end{abstract}
\section{Introduction}
{T}{he} Type-III Universal Number System (unum), commonly referred to as the \textit{posit} number system \cite{ref1}, represents a modern alternative to the IEEE Standard 754 Floating-Point (\textit{float} here, for short) format \cite{ref2}. Posits were introduced to address inefficiencies in float arithmetic by providing tapered relative accuracy (greater relative accuracy for magnitudes close to unity, and less for extremely large or small magnitudes), resulting in improved numerical accuracy and energy efficiency in many workloads \cite{ref3}.

A posit is defined primarily by two parameters: its word size $N$ and its exponent size \textit{ES}. Following the ratified posit standard \cite{ref12}, the \textit{ES} is set to 2, which offers a balance between precision and dynamic range. Each posit consists of four fields: a sign bit, a regime field, an exponent field, and a fraction field. The run-length encoded regime compactly represents the scale of the more commonly used numbers, while the exponent and fraction together determine its fine-grained magnitude. With $s$ denoting the sign bit, $k$ the regime value, $e$ the exponent value, and $f$ the fraction, the numerical value of a posit can be expressed as
\begin{equation}
    (1 - 3s + f) \times 2^{(1-2s)\times(2^{\textit{ES}} \cdot k + e +s)}.
\end{equation}
This compact and self-normalizing structure eliminates special cases such as subnormals, infinities, and NaNs found in IEEE 754, with only two exceptions: zero and “Not a Real” (NaR).

Despite these strengths, posits face a key barrier to widespread hardware adoption: their variable-length field encoding and decoding logic creates higher complexity than float units of comparable precision \cite{ref4}. Consequently, hardware implementations must carefully optimize these stages to mitigate power, delay, and area overheads. 

Multiplication is a dominant operation in numerical computation, from image and signal processing to neural network inference. However, it is also among the most resource-intensive in terms of chip area, power, and latency. Division, while less frequent, remains computationally expensive and often dictates system throughput due to its inherently serial nature. In binary arithmetic, division algorithms such as digit recurrence, functional iteration, and table look-up, trade off area, delay, and precision in distinct ways \cite{ref5, ref6, ref7}.

Recent posit division designs \cite{ref8, ref9, ref10} largely mirror float techniques, highlighting the structural similarities between the two number systems. In this work, we adopt an approximate reciprocal-based division approach, inspired by Moroz et al. \cite{ref11}, and integrate it into a unified posit multiply-divide unit. This architecture performs exact multiplication and approximate division within the same circuit, achieving high accuracy and significant energy savings.

\begin{flushleft}
The main contributions of this work are as follows:
\end{flushleft}
\begin{itemize}
    \item A unified hardware architecture that performs both exact posit multiplication and approximate division, employing a low-cost reciprocal estimation within the decoder.
    \item Optimized posit decoding and encoding architectures that emphasize parallel field extraction and recombination, reducing power and delay significantly.
    \item A LUT-assisted approximate reciprocal computation that is fast and energy-efficient but may produce divisions that differ from a correctly-rounded reciprocal).
    \item A power-efficient significand multiplier featuring optimized partial product generation and reduction schemes.
\end{itemize}

\section{Related Work}
This work proposes a unified multiply–divide unit for posit arithmetic. We review previous designs for posit multiplication and division units that form the foundation of our architecture.

\subsection{PACoGen: Posit Arithmetic Core Generator}
PACoGen \cite{ref8} introduces a parameterized posit divider capable of supporting a variety of posit sizes and configurations. The main design parameters include word size ($N$), exponent size ($eS$), Newton–Raphson iteration count (NR$\_$Iter) and the bit widths of the lookup table (LUT) entries (IW$\_$MAX) and addresses (AW$\_$MAX). The divider features custom posit decoding, fraction division, and result packing modules.

The most resource-intensive step is fraction division, which computes the reciprocal of the denominator significand using a LUT-based seed followed by iterative refinement. The number of Newton–Raphson (NR) iterations doubles the precision with each iteration, as follows:
\begin{equation}
X_{i+1} = X_i \times (2 - D \times X_i)
\end{equation}
where $X_i$ denotes the reciprocal estimate in iteration $i$, and $D$ is the denominator fraction. Thus, each iteration entails two multiplications and one subtraction.

For 8-bit posits, LUT-stored seeds (9-bit accuracy) suffice without iterations. A 16-bit posit requires one NR iteration, while 32- and 64-bit formats need two and three iterations, respectively, reflecting the exponential growth of the maximum number of bits in the significand. Although more iterations improve accuracy, they increase delay and power. Once convergence is achieved, the final reciprocal, rounded, is multiplied by the numerator fraction and again rounded to obtain the quotient. The two rounding errors in computing $x\times (1/y)$ can differ from a singly-rounded quotient $(x/y)$ by $-1$, $0$, or $1$ Unit in the Last Place (ULP), placing the $x\times (1/y)$ approach in the ``approximate divider'' category.

\subsection{Posit Division Using the Newton–Raphson Method}
The design in \cite{ref23} follows an approach based on NR similar to \cite{ref8} but replaces the reciprocal seed derived from LUT with an estimate obtained through a 2's complement transformation of the denominator (with preserved sign bit). The subsequent NR iterations refine this initial estimate until the reciprocal is sufficiently precise. The required iteration count depends on the quality of the initial seed: Closer approximations reduce NR steps and consequently hardware latency and power.

\subsection{PLAUs: Posit Logarithmic Approximate Units}
Building on earlier logarithmic posit multipliers \cite{ref25}, the authors in \cite{ref24} extend the design to division and square root operations. Using Mitchell’s logarithmic approximation algorithm \cite{ref26}, the inputs are transformed from linear to logarithmic domain, where multiplication and division simplify to addition and subtraction. Mitchell’s method approximates $\log_2(x)$ by $x$ for $x \in [0,1)$, greatly simplifying hardware but introducing notable approximation error. Comparative analyzes in \cite{ref27} confirm that logarithmic posit units achieve much lower delay, power, and area than exact designs based on NR or non-restoring, although with significantly higher computational error that cannot be easily iteratively corrected.

Recently, logarithmic posit encodings have also been explored for DNN inference efficiency. Ramachandran proposed a distribution-aware logarithmic-posit co-design that achieves high efficiency–accuracy trade-offs, further motivating approximate arithmetic exploration within the posit domain \cite{ref23}.

\subsection{Parameterized Fused Division and Square Root Posit Architecture}
The design in \cite{ref9} presents a unified unit capable of performing both posit division and square root operations. It leverages shared logic between the non-restoring division and square root algorithms to minimize area. However, the non-restoring method remains relatively slow, since it generates only one quotient bit per cycle. For example, a 16-bit posit, comprising up to 11 fraction bits plus one implied bit, requires 12 cycles to produce a complete quotient.

\subsection{Design of a Power-Efficient Posit Multiplier}
Zhang and Ko \cite{ref13} propose a Booth-based post-transfer multiplier that integrates a decoder and a significand multiplier with power-optimized control logic. The stage of partial product generation (PPG) is divided into 16 regions based on input size. Since not all fraction bits contribute to the computation (depending on the regime size), unused bits are padded with zeros and gated to prevent unnecessary switching.

The control logic employs four vertical and four horizontal control signals to selectively activate multiplier regions, effectively reducing dynamic power. Each vertical signal controls four bits, and each horizontal signal controls two rows of partial products. Disabling inactive regions minimizes toggling, achieving significant energy savings without impacting computational accuracy.

\section{Proposed Unified Multiply-Divide Architecture}

\begin{figure*}[h!]
  \centering
  \captionsetup{justification=centering} 
  \includegraphics[width=1\linewidth]{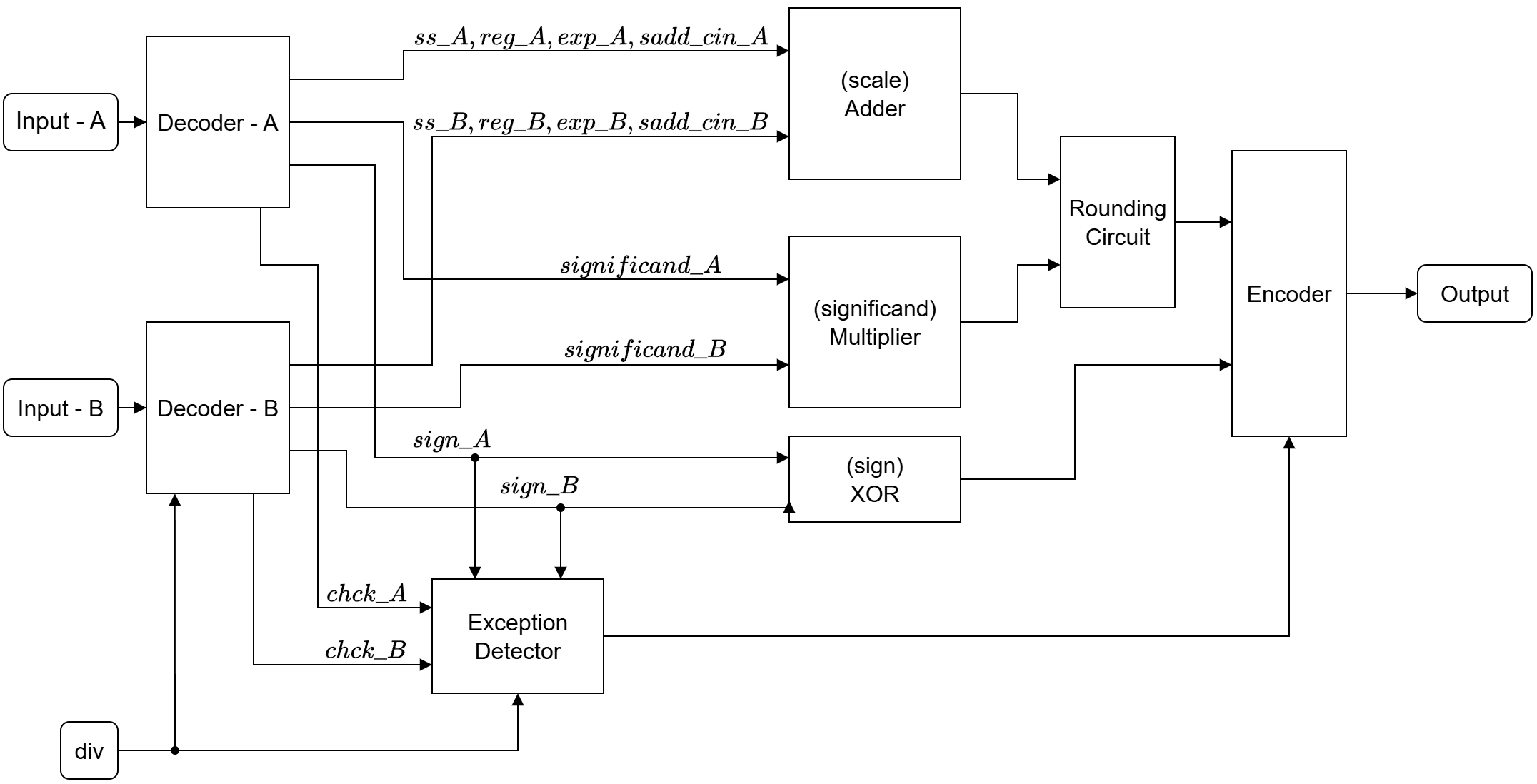}
  \caption{Proposed Hardware Architecture for Unified Multiply-Divide Unit}
  \label{fig:architecture_overview}
\end{figure*}

The proposed unified multiply-divide unit, shown in Fig.~\ref{fig:architecture_overview}, can perform either an exact multiplication or an approximate division depending on the mode of operation, with only a minor overhead in accuracy and delay for the latter. In the proposed method, division is implemented by computing the reciprocal of the divisor and multiplying it by the dividend.

The first step of all posit arithmetic operations is the decoding of inputs into their respective regime, exponent, and fraction fields, which can then be processed as needed. In hardware, the decoding process also involves performing a 2’s complement when the sign bit is set to \texttt{1}. This simplifies the formula for a posit value to

\begin{equation}
    (1 + f) \times 2^{(2^{\textit{ES}} \cdot k + e)}
\end{equation}

To minimize latency and hardware overhead, reciprocal functionality is directly integrated into the decoder within the presented architecture.

As shown in Fig.~\ref{fig:architecture_overview}, decoders take two posit numbers as inputs and output their sign bits, exponents, regimes, and fractions. When either input has fewer bits than the maximum supported configuration, the unused bits are padded with zeros. Because only one operand requires reciprocal computation, the design of the two decoders is intentionally asymmetric. For clarity, the decoder without reciprocal functionality is referred to as \textit{Decoder A}, and the one incorporating it as \textit{Decoder B}. Decoder B also accepts an additional control bit, \textit{div}, that determines whether to compute a standard or reciprocal output.

After decoding, the final output sign bit is determined by an XOR of the operand sign bits. The extracted regime and exponent values (\textit{reg\_A}, \textit{reg\_B}, \textit{exp\_A}, and \textit{exp\_B}) are combined to form the output scale (final exponent) in binary 2’s complement representation, simplifying the later encoding step. The significand fields from both decoders are multiplied using conventional methods, followed by rounding. The \textit{Scale Adder} also receives control and carry-in signals (\textit{sadd\_cin}) that ensure correct scaling when the fraction field is zero or when 2’s complementing occurs. Finally, the decoders generate exception-checking flags (\textit{chck}) for zero and NaR detection.

\subsection{Decoder A}

Decoder A extracts posit components and serves as the baseline for Decoder B. Functionally, it resembles previous implementations \cite{ref8,ref13,ref14} but with several structural optimizations. The design incorporates early partial processing before the 2’s complement step, using an intermediate control signal \textit{ctrl} derived from XORing the sign bit and the most significant regime bit. This signal also indicates the contribution of the sign for the final exponent and scale computation.

\begin{figure*}[h!]
  \centering
  \captionsetup{justification=centering} 
  \includegraphics[width=1.0\linewidth]{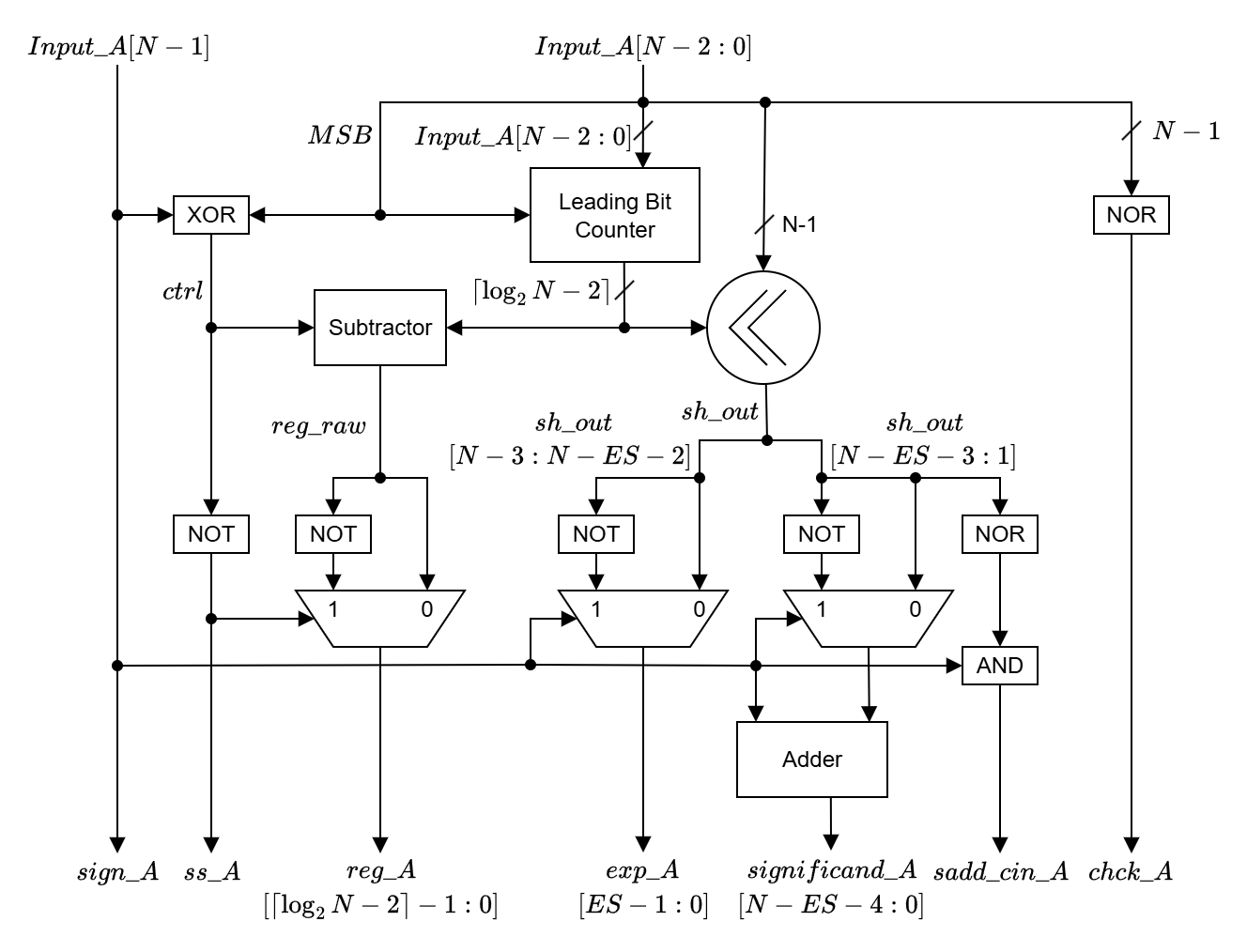}
  \caption{General Design of Proposed $\langle N,\textit{ES}\rangle$ Decoder A}
  \label{fig:decoderA}
\end{figure*}

Exception cases (all \texttt{0} bits for zero and all \texttt{0} bits except the sign bit for NaR) are detected, and the corresponding flag \textit{chck\_A} is asserted.  
The regime field is processed using a leading-bit counter (LBC), adapted from \cite{ref15}, to determine the terminating bit position. Depending on the polarity of the regime field, the LBC functions as a leading-\texttt{1} or leading-\texttt{0} counter. The magnitude of the derived regime $|k|$ is obtained by subtracting \textit{ctrl} from the count result, with the final sign applied based on \textit{ctrl}. In parallel, the exponent and fraction fields are shifted, inverted, and complemented as required. A reduction NOR of the preprocessed fraction bits generates \textit{sadd\_cin\_A}, used later during scale addition. This approach reduces the adder width and enhances the pipeline overlap between exponent and fraction calculations, improving the delay performance.

\subsection{Decoder B and Reciprocal Computation}

Decoder B extends Decoder A by introducing a single-bit input, \textit{div}, that enables the reciprocal calculation mode. When $\textit{div}=\texttt{0}$, the decoder behaves identically to Decoder A; when $\textit{div}=\texttt{1}$, it executes the proposed reciprocal algorithm.

\begin{figure}[h!]
  \centering
  \subfloat[Error \% vs. fraction at FB = 4]{%
    \includegraphics[width=0.45\linewidth]{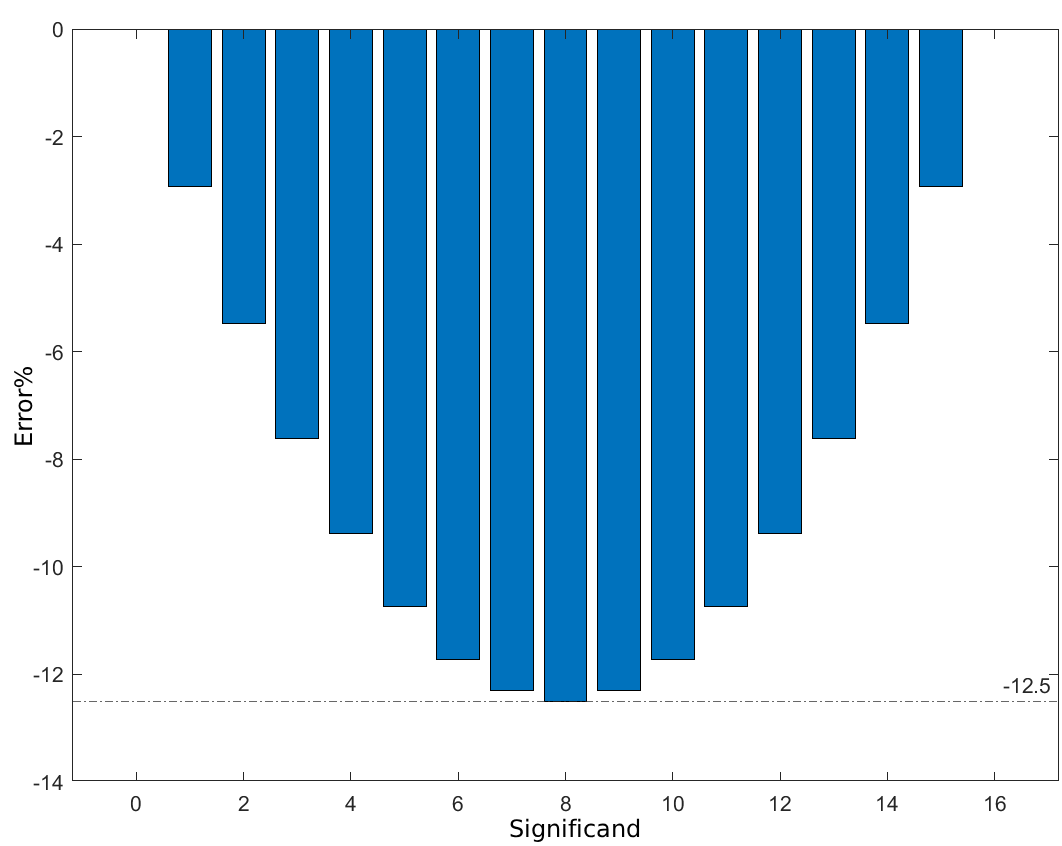}%
    \label{fig:error_fb4}}
  \hfill
  \subfloat[Error \% vs. fraction at FB = 11]{%
    \includegraphics[width=0.45\linewidth]{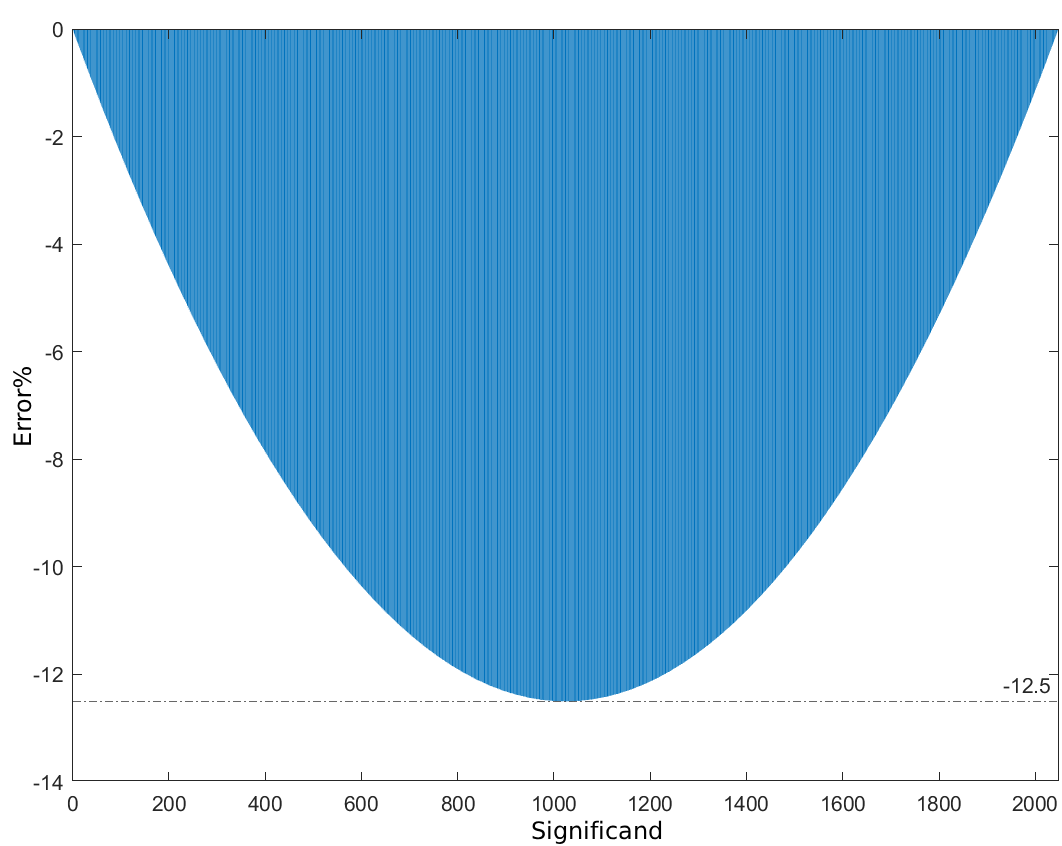}%
    \label{fig:error_fb11}}
  \caption{Error \% with respect to fraction for different fraction bit lengths FB.}
  \label{fig:error_fraction}
\end{figure}

When a posit number is 2’s complemented while keeping its sign bit constant, the resulting bit pattern approximates the reciprocal of the original value. For inputs with fraction $f=0$, this approximation is exact; for others, a small deviation exists because this approximates the hyperbolic curve $1/(1+f)$ by a line $1-f/2$, $0\leq f<1$. Since we operate on pure binary numbers, the fraction term effectively represents the smallest distinguishable interval within the given bit precision. Therefore, rounding this term to the nearest integer is equivalent to rounding to the nearest representable posit value. The proposed correction method adjusts this fraction using a precomputed error correction (EC) term to further minimize the approximation error.

To derive the appropriate EC value, we begin by defining the error as the difference between the true reciprocal of a posit number and the value obtained by 2’s complementing the input while keeping the sign bit constant. Denoting the original posit value as
\[
X = x(1 + f),
\]
where $x$ represents the scaling factor contributed by the regime and exponent fields, and $f$ is the fraction. Taking the 2’s complement yields the approximate reciprocal:
\[
\Bar{X} = \frac{1}{2x}(2 - f).
\]
\noindent Introducing an error correction term $EC$ (scaled to fraction bits FB) gives the following:
\[
\Bar{X} = \frac{1}{2x}\left(2 - f - \frac{EC}{2^{\text{FB}}}\right).
\]\

\noindent The exact reciprocal of $X$ is:
\[
\frac{1}{X} = \frac{1}{x(1+f)}.
\]\

\noindent Hence, the relative error becomes:
\[
\mathrm{error}_\mathrm{frac} = \frac{\text{Actual} - \text{Approximate}}{\text{Actual}} = 
-\frac{f(1-f)}{2} + \frac{EC(1+f)}{2^{(\text{FB}+1)}}.
\]\

\noindent To minimize the error, we differentiate $(\mathrm{error}_\mathrm{frac})^2$ with respect to $EC$ and set the derivative to zero:
\[
\frac{d}{d(EC)}(\mathrm{error}_\mathrm{frac})^2 = 0 \Rightarrow 
EC = \frac{f(1-f)}{(1+f)} 2^{\text{FB}}.
\]

Since the hardware implementation requires integer correction values, $EC$ is rounded to the nearest integer. This rounding effectively corresponds to selecting the nearest representable number within the precision bounds of the posit format.

Precomputed EC values are stored for specific fraction intervals in a look-up table (LUT). Intermediate values are corrected using the nearest available EC entry, and the accuracy improves as more sampling points are stored.

Table 1 summarizes the accuracy improvement achieved through the EC-based correction for various posit configurations.

\begin{table}[!ht]
\renewcommand{\arraystretch}{1.25}
\caption{Accuracy and LUT size versus different posit configurations}
\label{table_accuracy_lut}
\centering
\begin{tabular}{|c|c|c|}
\hline
\textbf{\begin{tabular}[c]{@{}c@{}}Posit Configuration \\ ($N$, \textit{ES}, MSB used)\end{tabular}} & \textbf{\begin{tabular}[c]{@{}c@{}}Accuracy\\ $\left| (Avg. Error\% )\right|$\end{tabular}} & \textbf{LUT Size} \\ \hline
16, 2, 0                                                                                    & 8.3333                                                                                          & -                 \\ \hline
16, 2, 5                                                                                    & 0.3834                                                                                          & 0.036 KB          \\ \hline
16, 2, 8                                                                                    & 0.0434                                                                                          & 0.288 KB          \\ \hline
32, 2, 0                                                                                    & 8.3333                                                                                          & -                 \\ \hline
32, 2, 5                                                                                    & 0.3932                                                                                          & 0.100 KB          \\ \hline
32, 2, 8                                                                                    & 0.0488                                                                                          & 0.800 KB          \\ \hline
\end{tabular}
\end{table}

This approach simplifies the reciprocal calculation into three steps:\\
\begin{enumerate}
\item{Apply 2's complement to the posit input while keeping the sign bit constant when necessary.}
\item{Use the most significant fraction bits to determine the address of the corresponding EC value in the LUT.}
\item{Subtract the EC value from the processed fraction.}
\end{enumerate}
\noindent
\newline
In conventional fast division algorithms \cite{ref16,ref17}, the seed is obtained from a LUT and a sequential process is performed to improve its relative accuracy. Whereas in the proposed method, the 2's complemented fraction directly acts as a seed, with the accuracy being improved in a simple combining operation, with an accuracy penalty.

\begin{figure*}[h!]
  \centering
  \captionsetup{justification=centering} 
  \includegraphics[width=1.0\linewidth]{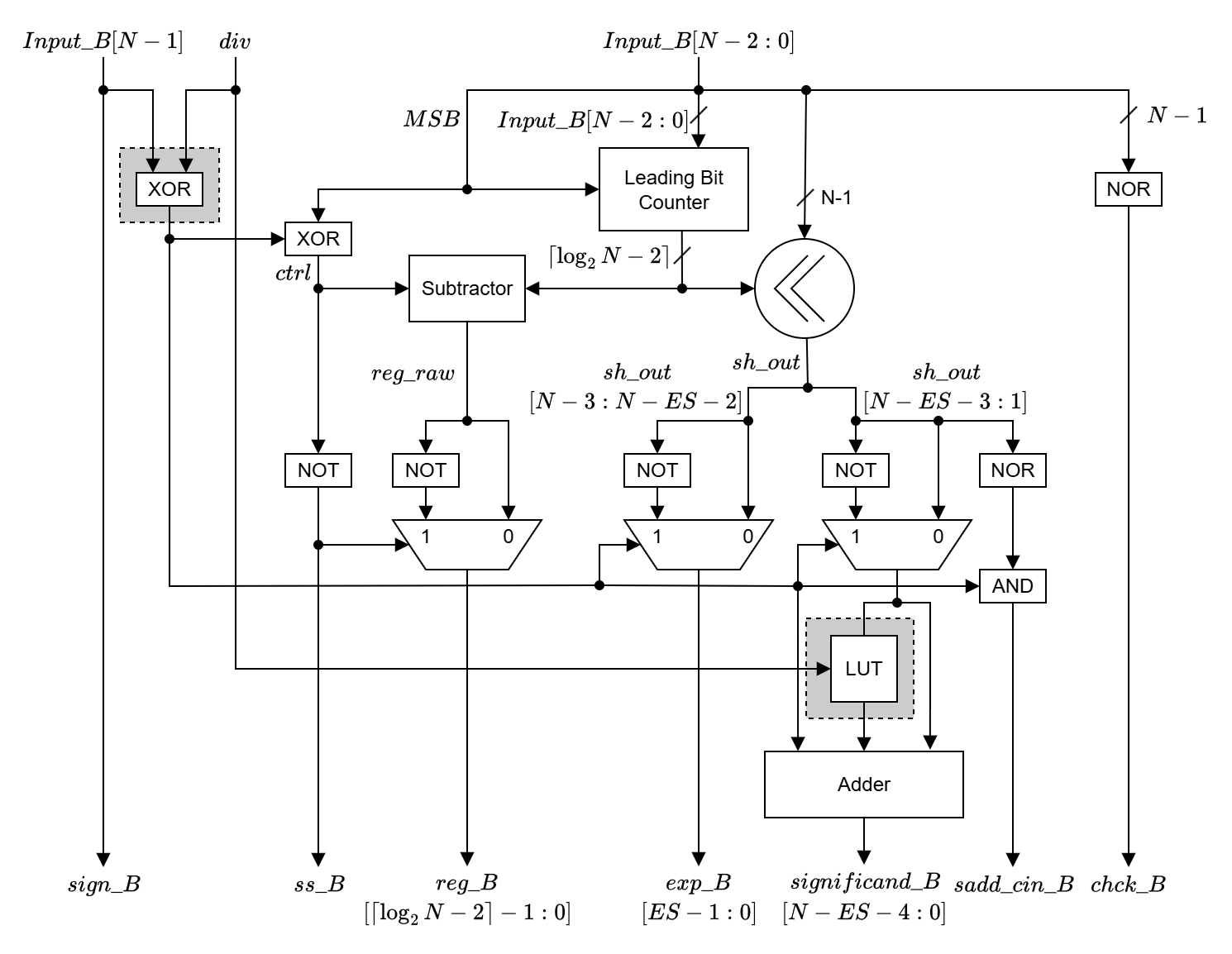}
  \caption{General Design of Proposed $\langle N,\textit{ES}\rangle$ Decoder B
  }
\end{figure*}

\subsubsection{Hardware Design}
Decoder B is a modified version of Decoder A with an additional XOR gate and a LUT, as highlighted in gray in Fig.4. The decoder accepts an additional single-bit input called \textit{div} and its functionality is identical to Decoder A when the \textit{div} bit is set to \texttt{0}. The input needs to be 2's complemented only when one of the sign bits or the \textit{div} bit is set to \texttt{1}. To achieve this, the two input bits are XORed to generate an internal control signal that takes the role of the sign bit in Decoder B for all common blocks between the two decoders.
\par
When the \textit{div} bit is set to \texttt{1}, the most significant bits of the preprocessed fraction are used to address the stored EC values in the LUT. The LUT address is 2's complemented when the sign bit is \texttt{1} since the LUT values are ordered for positive inputs. To ensure EC only comes into play when performing division, the LUT output is gated with the \textit{div} input. Since the EC values obtained are always subtracted from the fraction, the values stored in the LUT are complemented beforehand. 
\par

\subsection{Significand Multiplier}

The significand multiplier circuit efficiently processes inputs from Decoder A and Decoder B, performing a multiplication operation regardless of the mode of operation. An exact unsigned 12-bit Booth multiplier encoded with radix-8\cite{ref20} has been employed in the significand multiplication process for a $\langle16,2\rangle$ posit. The Booth multiplier has been modified in the partial product generation (PPG) stage to optimize the generation of negative partial products within the partial product array compared to a standard Booth process. This optimization significantly reduces dynamic power dissipation in the circuit. Additionally, the partial product reduction (PPR) stage has been trimmed to minimize both area and power dissipation. 
\par 

\subsubsection{Partial Product Generation}

In a general Booth multiplication process, each partial product row within the array is generated using Booth encoder circuits and PPG blocks. The encoder derives control signals from the multiplier operand bits, indicating the value by which the multiplicand should be multiplied to generate the partial product row. Some cases result in the partial product being a negative multiple of the multiplicand. In such instances, an XOR gate executes 1's complement operation on each partial product bit before sending it to the reduction structure. The sign bit is then placed in the LSB column of the following row within the partial product array, completing the 2's complement.
\par

The existing literature, such as \cite{ref13,ref18}, optimizes multiplier circuits to minimize dynamic power dissipation. Dedicated control logic blocks are proposed to prevent unnecessary switching during 2's complement operations. The input significand size ($\mathrm{sig}\_\mathrm{size} = N - \textit{ES} - \mathrm{regime}\_\mathrm{size}+1$) is managed by padding \texttt{0} bits to match the largest significand size when there is a smaller number of significand bits. The control circuit prevents the XOR gate switching of padded \texttt{0} bits, reducing dynamic power dissipation. However, additional logic is required to position the sign bit towards the LSB because of the varying number of non-zero significand bits.
\par 
\begin{figure}[b]
  \centering
  \captionsetup{justification=centering} 
  \includegraphics[width=0.55\textwidth]{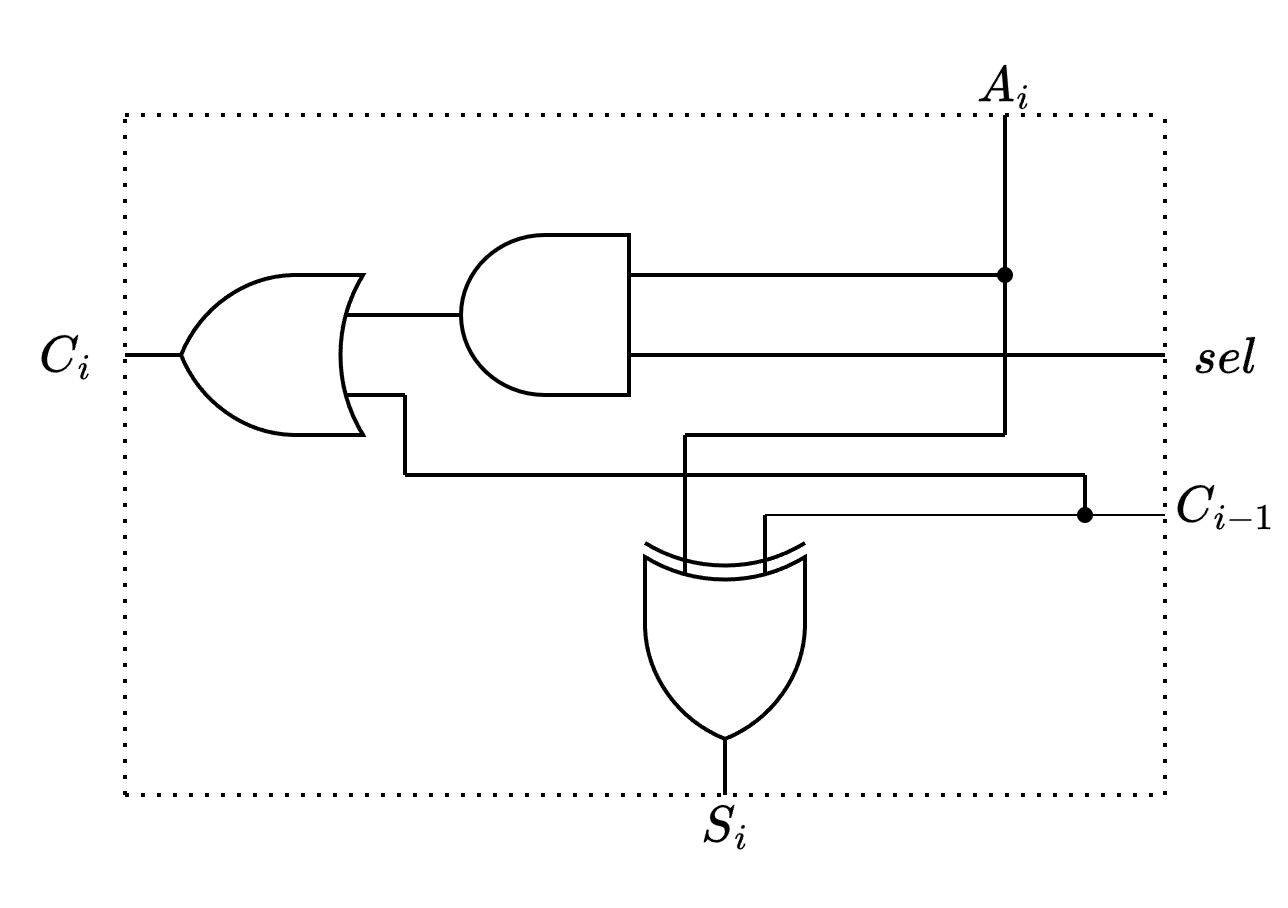}
  \caption{2's complement block used in PPG}
\end{figure}

\begin{figure*}[ht]
  \centering
  \captionsetup{justification=centering} 
  \includegraphics[width=0.98\textwidth]{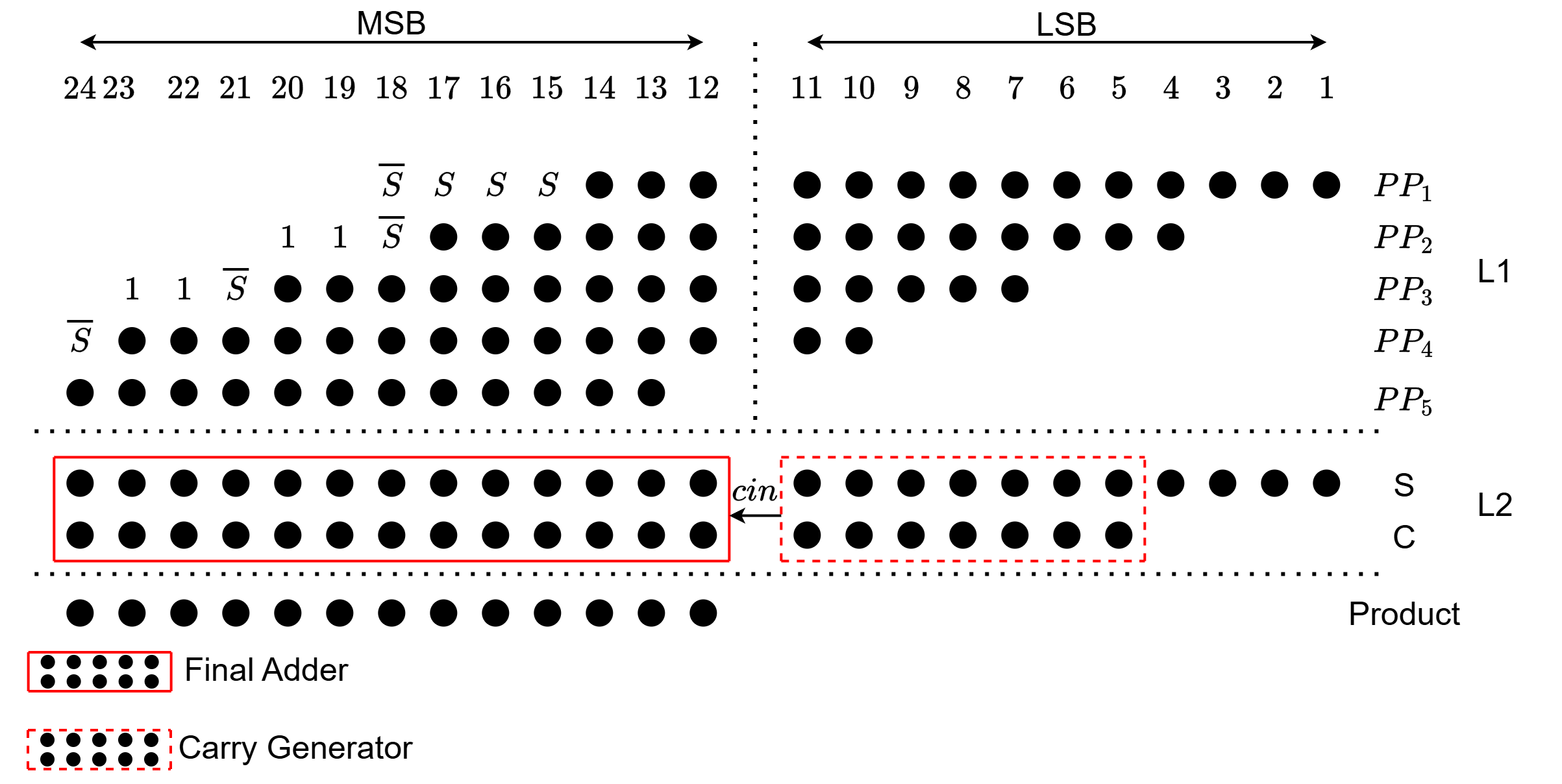}
  \caption{Partial product array of proposed significand multiplier}
\end{figure*}
The proposed multiplier design adopts a streamlined 2's complement methodology inspired by \cite{ref19}, eliminating the necessity for dedicated control circuits to prevent unwarranted bit switching. In each PPG block, the final XOR gate stage is omitted, and each bit in the partial product row before 2's complementing is directed to a dedicated 2's complement block shown in Fig.5. Unlike the traditional 1's complement operation followed by addition, the 2's complement circuit leaves \texttt{0} bits on the LSB side untouched until the first \texttt{1} bit is encountered, after which all remaining bits are inverted. As illustrated in Fig.5, an input bit $A_i$ undergoes inversion only when its corresponding $C_{i-1}$ bit is set to logic high and the \textit{sel} signal is asserted. The first $C_i$ signal to reach logic high requires the input bit of the previous stage ($A_{i-1}$) to be at logic high along with the \textit{sel} bit. Once any $C_i$ signal is set, all subsequent $C_{i+1}$ signals are activated, allowing bits $A_i$ onward to switch. This method ensures that until the leading \texttt{1} from the LSB is found, no bit inversion occurs, and switching begins from the next stage onward. This innovative block eliminates the need for a dedicated control circuit block, additional logic for sign bit positioning based on significand size in the reduction tree, and the inclusion of extra sign bits at the LSB of each row, simplifying the reduction structure when compared to existing designs.
\par

\subsubsection{Partial Product Reduction}
The partial product array of the proposed significand multiplier is depicted in Fig.6. Using radix-8 Booth encoding in a 12-bit unsigned multiplication process results in five rows of partial products. The entire reduction structure has been implemented in two stages: L1 and L2. In the L1 stage, a standard Wallace tree reduction is performed on the partial product array using 5:2 compressors, 4:2 compressors, full adders, and half adders. After this level of reduction, only two rows of bits ($S$ and $C$) remain for further reduction. The L2 stage employs a distinctive reduction approach by splitting the array into two distinctive MSB and LSB portions, which differs from the exact posit multipliers in the literature. Instead of calculating the entire 24-bit sum, only the 13 bits from the MSB side are computed, truncating the remaining bits with the product still being exact. This methodology is driven by the fact that a $\langle16,2\rangle$ posit has a maximum of 12 significand bits, including the implied bit. An extra product bit is computed for normalization and rounding, making the product size 13 bits. A 13-bit hybrid carry-select adder, combining Kogge Stone Adders (KSA) \cite{ref21} and Carry-Lookahead Adders (CLA) \cite{ref22}, serves as the final adder. It takes a carry from a 7-bit CLA-based carry generator, which generates only carry signals to save area compared to a full adder circuit. This reduction methodology contrasts with the existing literature, where a 24-bit addition process is typical. Additionally, the use of a 2's complement block in partial product generation ensures the absence of extra sign and zero bits on the LSB side of each row, simplifying the reduction process.

\begin{figure*}[ht]
  \centering
  \captionsetup{justification=centering} 
  \includegraphics[width=0.9\textwidth]{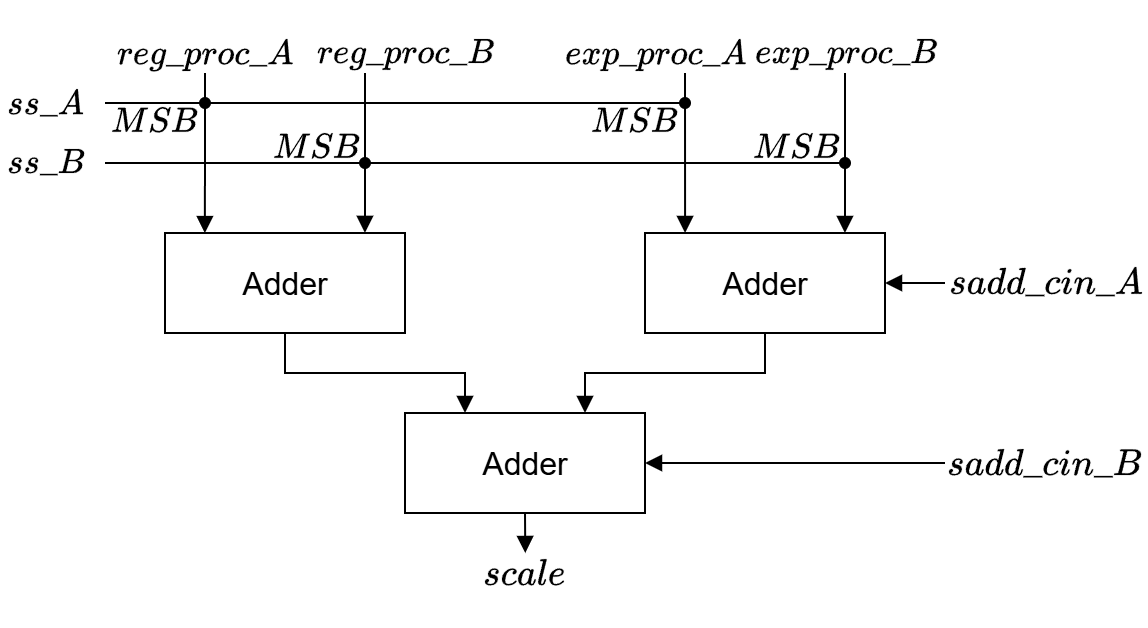}
  \caption{Proposed Scale Adder}
\end{figure*}

\subsection{Scale Adder}

The scale addition module, shown in Fig.7, receives as input the processed regimes, exponents, generated carry-in signals, and the final exponent signs from both decoders, culminating in the generation of the final exponent represented in 2's-complement form, called the \textit{scale}, facilitating the subsequent encoding process. 
\par 
The exponents and one of the generated carry-ins from the decoders are added together in a single step. Concurrently, the regimes are added in a separate adder. Finally, the results of the previous two additions along with the remaining previously generated carry-in are added to obtain the \textit{scale}, which is then sent to the rounding circuit. It should be noted that the regime sum is first left shifted by \textit{ES} places before being added to the exponent sum.

\subsection{Exception Detector}

If either of the inputs are zero or NaR, the usual data path cannot be used since they are exception cases in the posit number system. To account for the various combinations of exception cases, the exception detector module takes the \textit{chck} bits, \textit{sign} bits, and the \textit{div} bit to output a 2-bit control signal, \textit{excep}. The meaning of the different values the signal can take is highlighted in Table III. The \textit{excep} bits are later used to correct the output during the encoding stage.

\begin{table}[!ht]
\renewcommand{\arraystretch}{1.25}
\caption{Meaning of \textit{excep} bits}
\label{table_excep_bits}
\centering
\resizebox{\columnwidth}{!}{  
\begin{tabular}{|c|c|c|c|c|c|c|c|}
\hline
\textbf{\textit{div}} & \textbf{\textit{sign\_A}} & \textbf{\textit{chck\_A}} & \textbf{\textit{sign\_B}} & \textbf{\textit{chck\_B}} & \textbf{\textit{excep[1]}} & \textbf{\textit{excep[0]}} & \textbf{Output} \\ \hline
x            & x                & 0                & x                & 0                & 0                     & 0                     & normal          \\ \hline
0            & 0                & 1                & x                & 0                & 0                     & 1                     & zero            \\ \hline
0            & x                & 0                & 0                & 1                & 0                     & 1                     & zero            \\ \hline
0            & x                & 0                & 1                & 1                & 1                     & 1                     & NaR             \\ \hline
1            & x                & 0                & 1                & 1                & 0                     & 1                     & zero            \\ \hline
1            & 0                & 1                & 0                & 1                & 1                     & 1                     & NaR             \\ \hline
\end{tabular}
}
\end{table}

\begin{figure*}[!ht]
  \centering
  \captionsetup{justification=centering} 
  \includegraphics[width=0.9\linewidth]{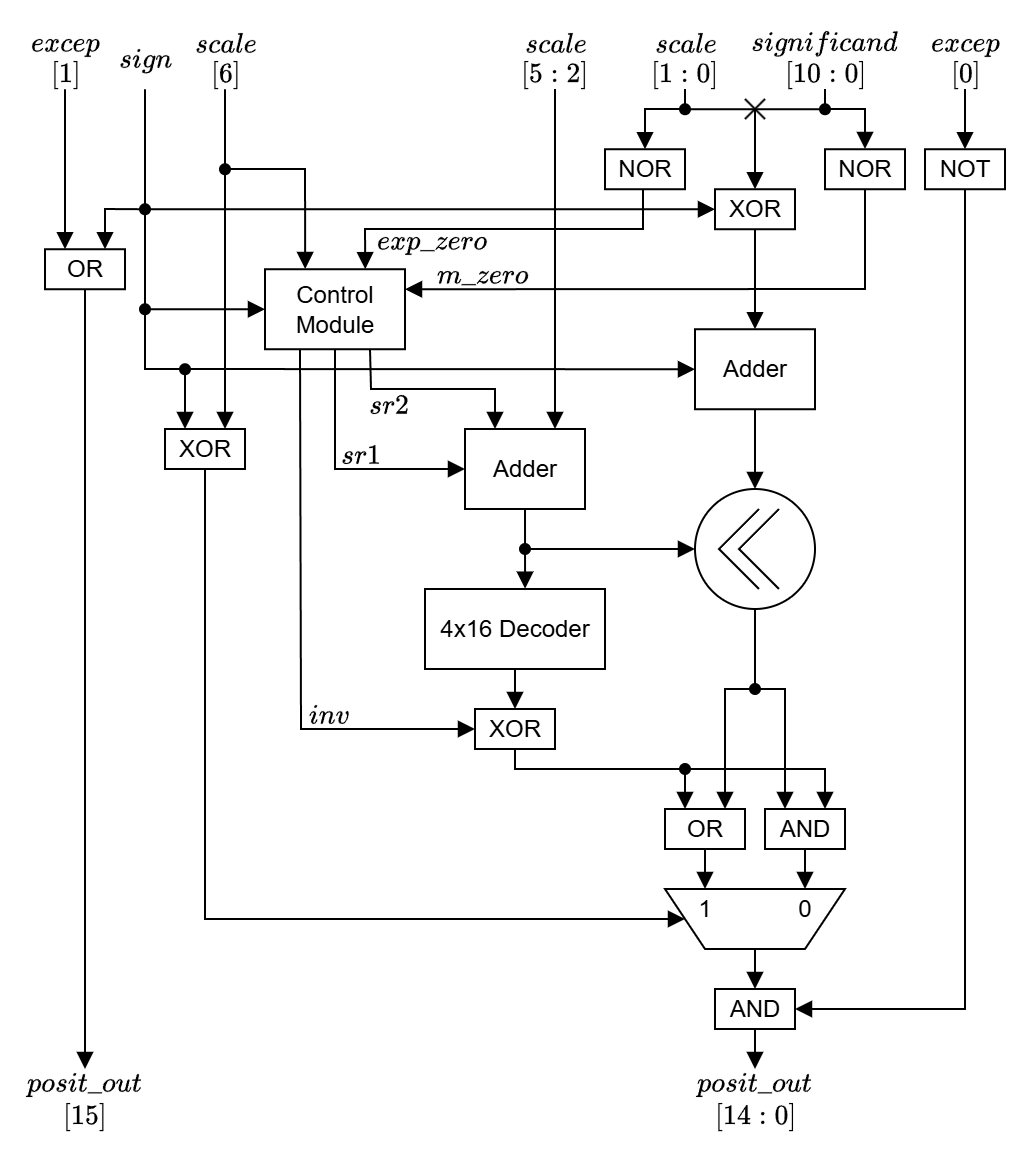}
  \caption{Proposed Encoder for $\langle16,2\rangle$ Posits}
\end{figure*}

\subsection{Encoder}

The Encoder Module, shown in Fig.8, serves a crucial role in posit arithmetic by consolidating the sign, the final exponent (\textit{scale}), and the fraction into the final posit output. Reflecting the design philosophy applied to decoders, this module prioritizes the parallelization of data processing. To allow for a change in the order of processing seen in existing designs \cite{ref8,ref14} a simple control module was created that outputs intermediary control signals \textit{sr1}, \textit{sr2} and \textit{inv}. 
\par
The \textit{sr1} signal accommodates scenarios where the output sign and the MSB of the regime are \texttt{01} or \texttt{10}, respectively, when the regime field is one position longer than implied by the total exponent. Another shift signal, \textit{sr2}, addresses cases where both the exponent and fraction fields are zero, and the output sign is \texttt{1}, necessitating a carry-in into the regime field. The two shift-right signals are combined with the processed regime obtained from the scale to determine the magnitude by which the exponent and fraction fields should be shifted. Subsequently, this amount of shift is sent to a $4\times16$ decoder to ensure the correct placement of the terminating one. In instances where the output sign and MSB of the regime are \texttt{01} or \texttt{10}, the output of the classical decoder must be inverted. This is required since the regime field in these cases needs to be repeating ones followed by a zero. As such, the output of the decoder undergoes inversion based on the \textit{inv} signal when necessary.
\par 
In parallel, the exponent and fraction fields are processed based on the sign bit. They are concatenated and shifted by the previously determined amount of shift. This shifted result is then subjected to a logical OR or AND operation with the output of XOR gates that follow the decoder, the choice being governed by the \textit{inv} signal. Finally, the sign bit is concatenated with the result obtained to output the encoded posit. 
\par
If the \textit{excep}[0] bit is set, all output position fields, except the sign field, are forced to \texttt{0}s. Moreover, the sign field undergoes a logical OR operation with \textit{excep}[1] to address exception cases.

\section{Results}
This section presents both the accuracy and hardware synthesis results of the proposed posit multiply–divide unit. The accuracy analysis focuses on the error characteristics of the reciprocal approximation, while the synthesis results compare the proposed design with existing implementations in terms of power, area, delay, and power-delay product (PDP).

\subsection{Accuracy Results}
The proposed division algorithm computes an approximate reciprocal followed by an exact multiplication. Hence, any inaccuracy originates solely from the reciprocal approximation. The accuracy was evaluated using three standard metrics --- mean error distance (MED), mean relative error distance (MRED), and normalized MED (NMED) --- for all $2048$ possible fraction values in $\langle16,2\rangle$ posits. The results are summarized in Table~\ref{table_division_error_metrics}.

\begin{table}[!h]
\renewcommand{\arraystretch}{1.25}
\caption{Proposed Division Error Metrics Analysis}
\label{table_division_error_metrics}
\centering
\begin{tabular}{|c|c|c|c|}
\hline
\textbf{LUT Size} & \textbf{MED (\%)} & \textbf{MRED (\%)} & \textbf{NMED (10$^{-3}$)} \\ \hline
$2^5 \times 9$     & 0.2677           & 0.3834            & 0.1850                     \\ \hline
$2^6 \times 9$     & 0.1322           & 0.1800            & 0.1869                     \\ \hline
$2^7 \times 9$     & 0.0656           & 0.0902            & 0.1816                     \\ \hline
$2^8 \times 9$     & 0.0332           & 0.0434            & 0.1910                     \\ \hline
\end{tabular}
\end{table}

Doubling the LUT size approximately halves the MED, thereby improving accuracy, though at the cost of increased circuit area. The proposed design employs a $2^5 \times 9$ LUT, which offers a practical balance between accuracy and resource usage.

The proposed decoder computes an approximate reciprocal similar to PACoGen~\cite{ref8}, but directly uses this approximation during multiplication rather than refining it through Newton–Raphson (NR) iterations. The experimental results show that the proposed design achieves a more accurate reciprocal seed (MRED = 0.38\%) than PACoGen (MRED = 0.61\%), despite using an $8\times$ smaller LUT. If both used an identical $2^8 \times 9$ LUT, the MRED of the proposed design would drop to just 0.0434\%, $14\times$ more accurate than PACoGen. This higher intrinsic accuracy suggests that the proposed approach could require fewer NR iterations for larger posit sizes, reducing hardware complexity in future extensions.

\subsection{Hardware Synthesis Results}
The proposed and reference designs were synthesized using the Cadence RTL Compiler v7.1 with a TSMC 90\,nm (slow-normal) process library. Power, area, and delay metrics were collected for all major components and compared with equivalent circuits in the existing literature. The following designs were used as benchmarks:
\begin{itemize}
    \item PACoGen~\cite{ref8}: Full divider data path (decode to encode).
    \item NPFD~\cite{ref9}: Significand division via non-restoring algorithm.
    \item POSIT-VAR~\cite{ref14}: Decoder, inexact multiplier, and encoder design.
    \item PEfPM~\cite{ref13}: Decoder and exact significand multiplier.
    \item Booth~\cite{ref20}: Exact significand multiplier circuit.
\end{itemize}

The functional blocks of each design were analyzed separately for a fair comparison.

\subsubsection*{Decoder Performance}
The proposed unit includes two decoder circuits, Decoder~A and Decoder~B. Decoder~A serves as the primary posit decoder, decomposing the input into sign, regime, exponent, and fraction fields. It achieves the lowest delay (2.464\,ns), outperforming the best existing design by 29.96\%, while also consuming less power and area. The overall PDP is 31.5\% lower than the next-best decoder and 61.4\% lower than PACoGen, as shown in Table~\ref{tab:decoder_comparison}.

\begin{table}[!h]
\renewcommand{\arraystretch}{1.25}
\caption{Comparison of proposed and existing decoder circuits}
\label{tab:decoder_comparison}
\centering
\normalsize
\resizebox{0.95\columnwidth}{!}{
\begin{tabular}{|c|c|c|c|c|c|c|c|}
\hline
\textbf{Design} & \textbf{Area ($\mu$m$^2$)} & \textbf{Area Ratio} & \textbf{Power ($\mu$W)} & \textbf{Power Ratio} & \textbf{Delay (ns)} & \textbf{Delay Ratio} & \textbf{PDP Ratio} \\
\hline
PACoGen~\cite{ref8} & 982 & 1 & 27.029 & 1 & 4.132 & 1 & 1 \\
PEfPM~\cite{ref13} & 851 & 0.87 & 22.123 & 0.82 & 3.719 & 0.9 & 0.74 \\
Posit-VAR~\cite{ref14} & 702 & 0.71 & 20.012 & 0.74 & 3.518 & 0.85 & 0.63 \\
Decoder A (\textit{div} = 0,1) & 603 & 0.61 & 17.568 & 0.65 & 2.464 & 0.59 & 0.38 \\
Decoder B (\textit{div} = 0) & 959 & 0.98 & 18.861 & 0.7 & 2.59 & 0.62 & 0.43 \\
Decoder B (\textit{div} = 1) & 959 & 0.98 & 32.588 & 1.21 & 4.584 & 1.1 & 1.3 \\
\hline
\end{tabular}}
\end{table}

Decoder~B handles reciprocal computation in division mode, adding an error-correction LUT that marginally increases area and power. In multiplication mode, this LUT is inactive, making its performance nearly identical to Decoder~A.

\subsubsection*{Significand Computation}
In the proposed design, both multiplication and division share the same significand computation unit. Division uses the approximate reciprocal from Decoder~B, while multiplication uses exact operands. Compared with NPFD and PACoGen, the proposed design eliminates iterative NR refinements and non-restoring loops, leading to substantial energy and delay savings. The results are shown in Table~\ref{tab:significand_comparison}.

\begin{table}[!h]
\renewcommand{\arraystretch}{1.5}
\caption{Comparison of significand computation processes}
\label{tab:significand_comparison}
\centering
\resizebox{\columnwidth}{!}{
\begin{tabular}{|c|c|c|c|c|c|c|c|c|}
\hline
\textbf{Operation} & \textbf{Design} & \textbf{Area ($\mu$m$^2$)} & \textbf{Area Ratio} & \textbf{Power ($\mu$W)} & \textbf{Power Ratio} & \textbf{Delay (ns)} & \textbf{Delay Ratio} & \textbf{PDP Ratio} \\
\hline
\multirow{4}{*}{Divide} 
 & PACoGen~\cite{ref8} & 6886 & 1 & 444.034 & 1 & 11.447 & 1 & 1 \\
 & PACoGen (no NR) & 2647 & 0.38 & 130.995 & 0.30 & 5.157 & 0.45 & 0.14 \\
 & NPFD~\cite{ref9} & 4216 & 0.61 & 226.806 & 0.51 & 22.148 & 1.93 & 0.98 \\
 & Proposed & 2597 & 0.37 & 115.792 & 0.26 & 4.846 & 0.42 & 0.11 \\ \hline
\multirow{3}{*}{Multiply} 
 & PEfPM~\cite{ref13} & 2771 & 1 & 145.678 & 1 & 4.461 & 1 & 1 \\
 & Booth~\cite{ref20} & 2546 & 0.92 & 159.321 & 1.09 & 4.412 & 0.99 & 1.08 \\
 & Proposed & 2597 & 0.94 & 115.792 & 0.79 & 4.846 & 1.09 & 0.86 \\
\hline
\end{tabular}}
\end{table}

The proposed significand unit achieves an 86\% reduction in PDP compared to existing divider architectures and up to 20\% lower power consumption compared to exact multipliers, while maintaining comparable area and delay.

\subsubsection*{Encoder Performance}
The proposed encoder emphasizes parallelization to minimize delay. As seen in Table~\ref{tab:encoder_comparison}, it achieves over 35\% lower delay, 23\% lower power, and 8\% smaller area than the best previous work (POSIT-VAR~\cite{ref14}).

\begin{table}[!h]
\renewcommand{\arraystretch}{1.25}
\caption{Comparison of encoder circuits in existing literature and the proposed design}
\label{tab:encoder_comparison}
\centering
\resizebox{\columnwidth}{!}{
\begin{tabular}{|c|c|c|c|c|c|c|c|}
\hline
\textbf{Design} & \textbf{Area ($\mu$m$^2$)} & \textbf{Area Ratio} & \textbf{Power ($\mu$W)} & \textbf{Power Ratio} & \textbf{Delay (ns)} & \textbf{Delay Ratio} & \textbf{PDP Ratio} \\
\hline
PACoGen~\cite{ref8} & 705 & 1 & 18.212 & 1 & 3.516 & 1 & 1 \\ 
Posit-VAR~\cite{ref14} & 601 & 0.85 & 16.516 & 0.9 & 3.052 & 0.87 & 0.78 \\ 
Proposed Encoder & 556 & 0.79 & 12.665 & 0.7 & 1.972 & 0.56 & 0.39 \\ 
\hline
\end{tabular}}
\end{table}

\subsubsection*{Full Datapath Comparison}
Finally, Table~\ref{tab:divider_comparison} compares the full datapath performance between the proposed architecture and PACoGen. An additional column labeled \textit{Exactness} clarifies whether the result produced is exact or approximate.

\begin{table}[!h]
\renewcommand{\arraystretch}{1.25}
\caption{Comparison between full posit divider datapaths}
\label{tab:divider_comparison}
\centering
\resizebox{\columnwidth}{!}{
\begin{tabular}{|c|c|c|c|c|c|c|c|c|}
\hline
\textbf{Design} & \textbf{Exactness} & \textbf{Area ($\mu$m$^2$)} & \textbf{Area Ratio} & \textbf{Power ($\mu$W)} & \textbf{Power Ratio} & \textbf{Delay (ns)} & \textbf{Delay Ratio} & \textbf{PDP Ratio} \\
\hline
PACoGen~\cite{ref8} & Exact & 10012 & 1 & 652.857 & 1 & 21.739 & 1 & 1 \\
PACoGen (no NR) & Approximate & 6872 & 0.69 & 288.516 & 0.44 & 17.474 & 0.80 & 0.35 \\
Proposed Multiplication & Exact & 5446 & 0.54 & 220.513 & 0.34 & 10.221 & 0.47 & 0.16 \\
Proposed Division & Approximate & 5446 & 0.54 & 254.148 & 0.39 & 12.828 & 0.59 & 0.23 \\
\hline
\end{tabular}}
\end{table}

The integrated results indicate that the proposed multiply–divide unit achieves a 77\% reduction in PDP and a 46\% reduction in area compared to PACoGen, with only a modest increase in power during the division mode due to reciprocal computation. Multiplication remains exact, while approximate division offers substantial efficiency gains with negligible accuracy loss. In general, the proposed architecture consistently outperforms existing designs in power, area, and delay, demonstrating its suitability for energy-efficient posit arithmetic hardware.

\section{Conclusion}

The proposed posit multiply-divide unit performs approximate division with minimal overhead over computing exact multiplication. The hybrid unit is on par with or better when compared to exact multipliers in the literature, while being significantly improved in all metrics against existing dividers, at the cost of accuracy. Efficient designs for all the common components present in posit arithmetic data paths were designed - decoders, significand multiplier, scale adder, and encoder. This allows for a sizable reduction in delay and power consumption compared to existing posit dividers and multipliers. A preliminary analysis also shows that the proposed design can easily be converted to an exact divider with a conventional NR method, while still retaining significant benefits in PDP. In the future, methods to increase the accuracy of the division process and the possibility of performing exact division using fewer hardware resources than existing literature will be explored.

%
%
%
\bibliographystyle{splncs04}
%

\end{document}